\documentclass{csmagclass}		
\usepackage{amsmath,amssymb}												
\begin{document}		
\headings																															 
\title{Anomalous thermodynamic response in the vicinity of a pseudo-transition of a spin-1/2 Ising diamond chain}
\author{Jozef Stre\v{c}ka\thanks{Corresponding author: jozef.strecka@upjs.sk}}
\affil{Faculty of Science of P. J. \v{S}af\'arik University, Park Angelinum 9, 04001 Ko\v{s}ice, Slovakia}
\maketitle

\begin{Abs}
The spin-1/2 Ising diamond chain in a magnetic field displays a remarkable pseudo-transition whenever it is driven sufficiently close to a ground-state phase boundary between a classical ferrimagnetic phase and a highly degenerate frustrated phase. The pseudo-transition of the spin-1/2 Ising diamond chain relates to intense thermal excitations from a nondegenerate ferrimagnetic ground state to a highly degenerate manifold of excited states with a frustrated character, which are responsible for an anomalous behavior of thermodynamic quantities. Temperature dependences of entropy and specific heat are indeed reminiscent of a temperature-driven phase transition of a discontinuous (entropy) or continuous (specific heat) nature though there are no true singularities of these thermodynamic quantities at a pseudo-critical temperature.
\end{Abs}
\keyword{Ising diamond chain, pseudo-transition, spin frustration}

\section{Introduction}

Recently, a few paradigmatic one-dimensional (1D) lattice-statistical models have aroused considerable research interest due to their anomalous thermodynamical behavior, which is strongly reminiscent of a temperature-driven phase transition \cite{gal15,str16,roj16}. However, the pseudo-transitions of 1D lattice-statistical models differ from true phase transitions in several important respects. The first-order derivatives of the free energy (e.g. entropy or magnetization) display near a pseudo-critical temperature a very steep but still continuous change instead of a true discontinuity, while the second-order derivatives of the free energy (e.g. specific heat or susceptibility) exhibit in the vicinity of a pseudo-critical temperature a very sharp but still finite peak instead of a true power-law divergence \cite{sou17}. In spite of this fundamental difference the pseudo-transitions of 1D lattice-statistical models are accompanied by a substantial rise of the correlation length near a pseudo-critical temperature \cite{car19} and, moreover, sizable peaks of the second-order derivatives of the free energy follow a striking power-law behavior characterized by a universal set of pseudo-critical exponents \cite{roj19}. 

In the present work, we will enrich realm of 1D lattice-statistical models displaying pseudo-critical behavior by a spin-1/2 Ising diamond chain, which has been marginally examined in Refs. \cite{can04,can06}. It is noteworthy, however, that a pseudo-transition of the spin-1/2 Ising diamond chain in a magnetic field has been completely overlooked in both earlier studies \cite{can04,can06} and thus remained unexplored yet.

\section{Spin-1/2 Ising diamond chain in a magnetic field}

\begin{figure}[t]
\begin{center}
\includegraphics[width=0.5\textwidth]{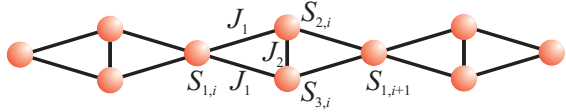}
\end{center}
\vspace{-0.6cm}
\caption{A schematic illustration of the spin-1/2 Ising diamond chain.}
\label{fig1}
\end{figure}

Let us consider the spin-1/2 Ising diamond chain in a magnetic field, which is schematically illustrated in Fig.~\ref{fig1} and mathematically defined through the Hamiltonian
\begin{eqnarray}
{\cal H} = J_2 \sum_{i=1}^N S_{2,i} S_{3,i} + J_1 \sum_{i=1}^N (S_{2,i} + S_{3,i}) (S_{1,i} + S_{1,i+1}) 
- h \sum_{i=1}^N \sum_{j=1}^3 S_{j,i}.
\label{eq:ham}
\end{eqnarray}
Here, $S_{j,i} = \pm 1/2$ are three Ising spin variables belonging to the $i$th unit cell, whereas the subscript $j=1$ ($j=2,3$) determines whether the relevant Ising spin variable is assigned to a monomeric (dimeric) site, respectively. Thus, the interaction parameter $J_1$ refers to the monomer-dimer coupling constant, $J_2$ is the intra-dimer coupling constant and the last term is the standard Zeeman's term associated with the magnetic field $h$. 

It is worthwhile to recall that the spin-1/2 Ising diamond chain has been previously exactly solved via a decoration-iteration transformation as a special limiting case of a more general spin-1/2 Ising-Heisenberg diamond chain \cite{can04,can06}. From this perspective, one may directly proceed to a discussion of the pseudo-critical behavior of the spin-1/2 Ising diamond chain in a magnetic field. Before doing this, however, let us recall three ground states emergent in the overall ground-state phase diagram (see Fig. 2 in Ref. \cite{can04}) of the spin-1/2 Ising diamond chain in a magnetic field: the ferrimagnetic phase $|\mbox{FRI} \rangle = \prod_i |\!-\!\frac12 \rangle_{1,i} |\frac12 \rangle_{2,i} |\frac12 \rangle_{3,i}$ is being the respective ground state in the parameter space $J_2<2J_1$ and $h<2J_1$, the frustrated phase $|\mbox{FRU} \rangle = \prod_i |\frac12 \rangle_{1,i} |\!\pm\! \frac12 \rangle_{2,i} |\!\mp\! \frac12 \rangle_{3,i}$ with a two-fold degeneracy of each antiferromagnetically aligned dimeric unit is being the respective ground state in the parameter region $J_2>2J_1$ and $h<J_1 + J_2/2$, and finally, the saturated paramagnetic phase $|\mbox{SPP} \rangle = \prod_i |\frac12 \rangle_{1,i} |\frac12 \rangle_{2,i} |\frac12 \rangle_{3,i}$ is being the respective ground state in the remaining parameter region.

\section{Pseudo-critical behavior}

\begin{figure}[t!]
\includegraphics[width=0.35\textwidth]{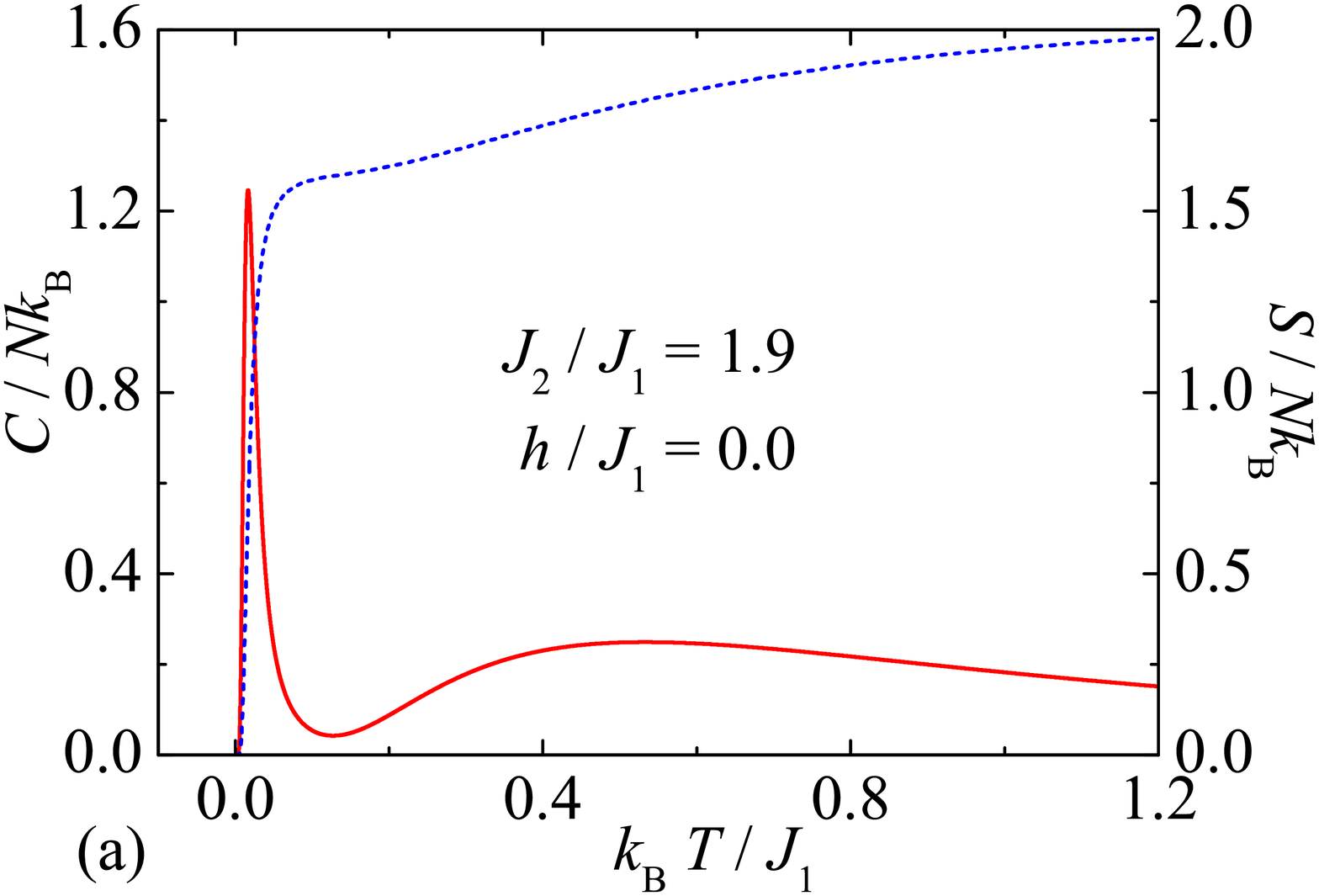}
\hspace{-0.55cm}
\includegraphics[width=0.35\textwidth]{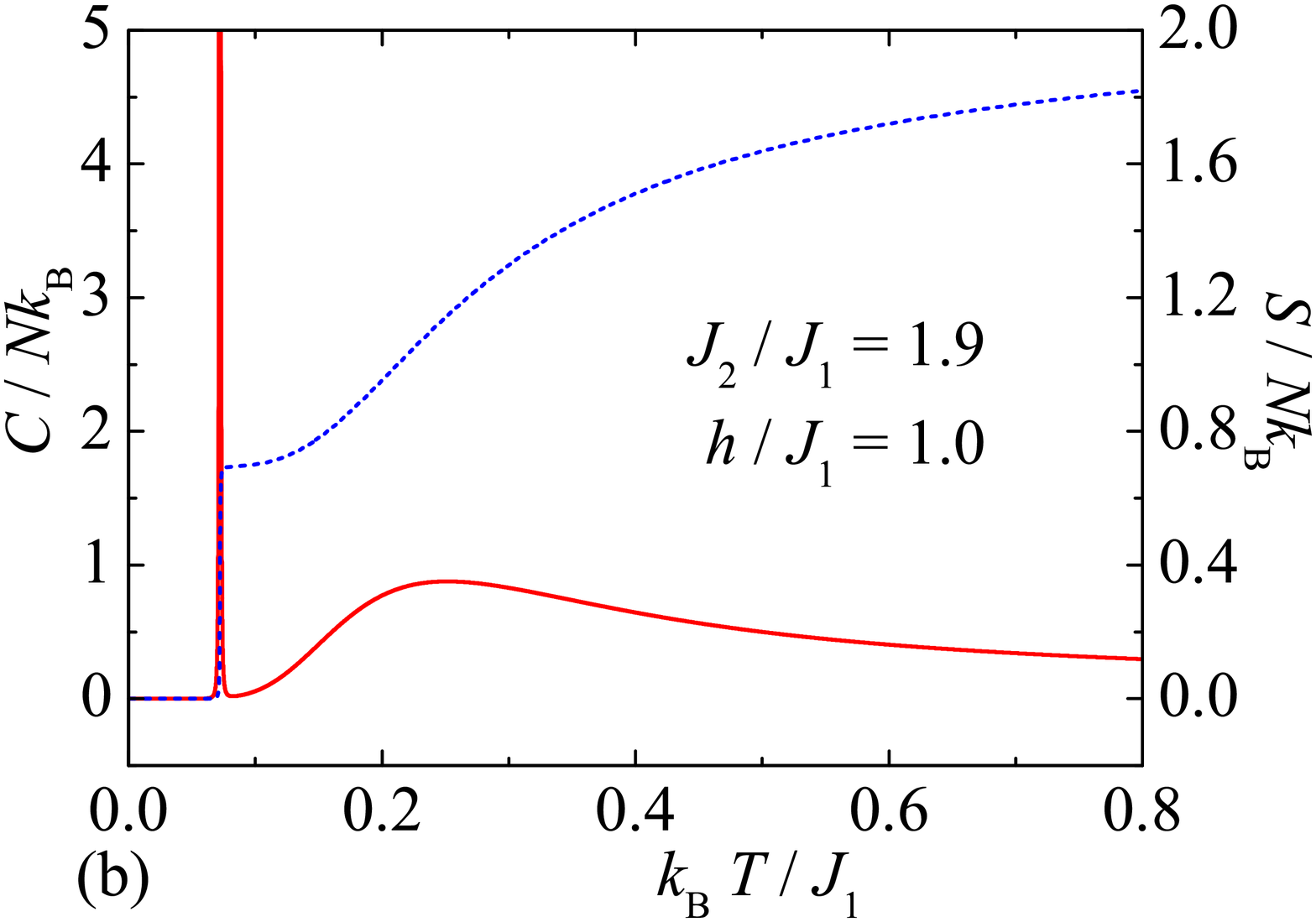}
\hspace{-0.55cm}
\includegraphics[width=0.35\textwidth]{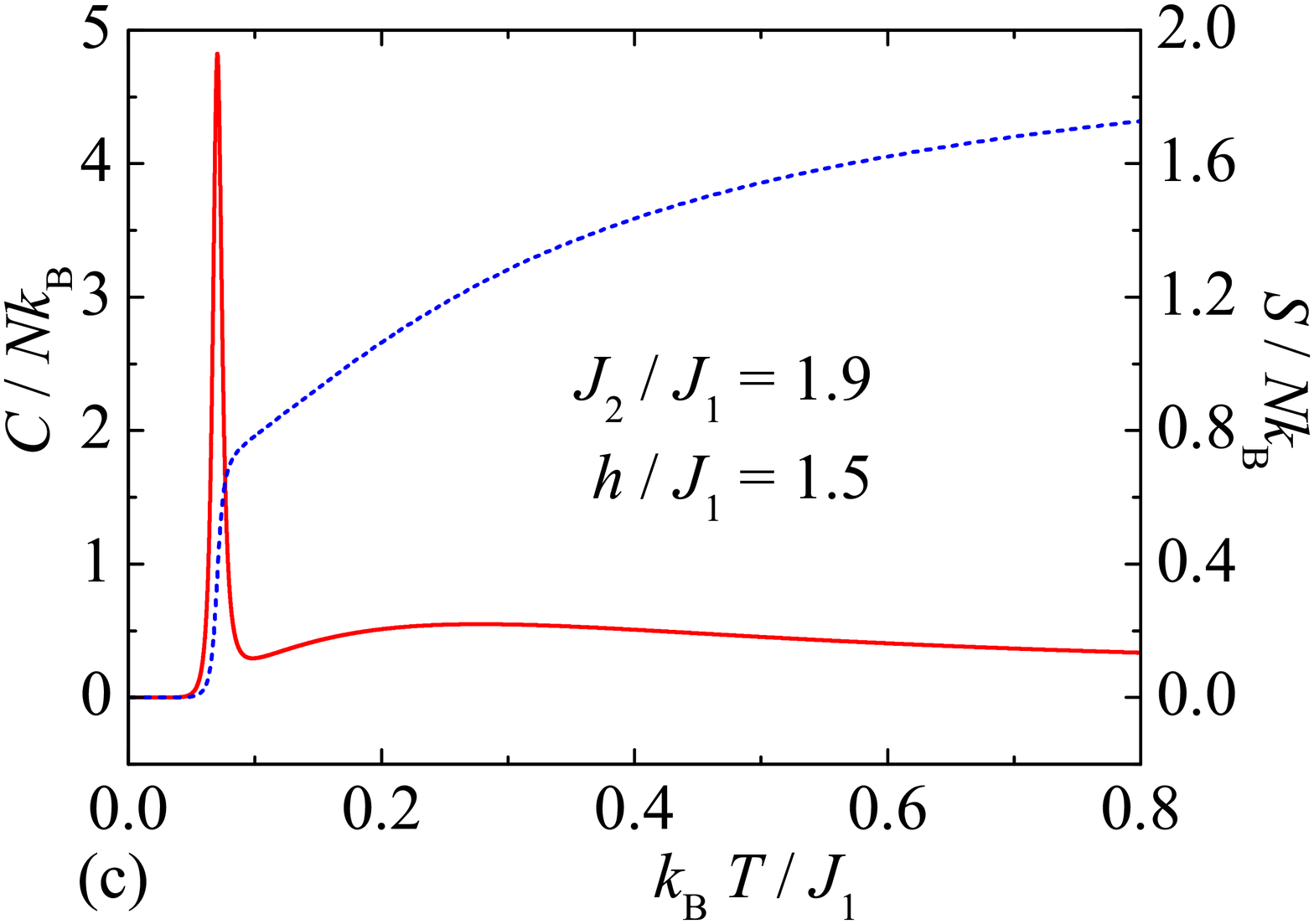}
\vspace{-0.7cm}
\caption{The specific heat (solid lines scaled to left axes) and entropy (broken lines scaled to right axes) for the fixed value of the interaction ratio $J_2/J_1 =  1.9$ and three different values of the magnetic field: (a) $h/J_1 = 0.0$; (b) $h/J_1 = 1.0$; (c) $h/J_1 = 1.5$.}
\label{fig2}
\end{figure}

The spin-1/2 Ising diamond chain displays a pseudo-transition only if a highly degenerate manifold of low-lying excited states is formed above the nondegenerate ferrimagnetic ground state due to an energy closeness to the macroscopically degenerate frustrated phase, which is ensured by suitable choice of the coupling ratio $J_2/J_1 \lesssim 2$ and small enough magnetic fields $h/J_1 < 2$. To support this statement, temperature variations of the entropy and specific heat are plotted in Fig. \ref{fig2} for the particular choice of the interaction ratio $J_2/J_1 = 1.9$ and three different values of the magnetic field. As one can see, the specific heat exhibits a temperature dependence with two well separated maxima, whereas sharper sizable peak emergent at a lower (pseudo-critical) temperature can be attributed to intense thermal excitations from the quasi-ferrimagnetic phase to the quasi-frustrated phase. The term 'quasi' is needed because of a lack of actual spontaneous long-range order at any finite temperature. In agreement with this picture, the spin-1/2 Ising diamond chain exhibits a pseudo-transition at the pseudo-critical temperature $k_{\rm B} T_p/J_1 = (2 - J_2/J_1)/\ln 4$, which can be obtained from a direct comparison of the free energies of the quasi-ferrimagnetic and quasi-frustrated phases when simply ignoring negligible changes of the internal energy and entropy at low enough temperatures (e.g. $k_{\rm B} T_p/J_1 \approx 0.072$ for $J_2/J_1 = 1.9$ at hand) \cite{gal15,str16}. Note furthermore that an abrupt change in the respective temperature dependence of the entropy followed by a quasi-plateau around $S \approx N k_{\rm B} \ln 2$ is indeed consistent with the residual entropy (macroscopic degeneracy) of the frustrated phase [see Fig. \ref{fig2}(b)-(c)], while the higher entropy $S \approx N k_{\rm B} \ln 4$ within a less evident quasi-plateau at zero magnetic field relates to an entropy gain arising from the monomeric spins [see Fig. \ref{fig2}(a)]. It is also quite obvious from Fig. \ref{fig2} that the most pronounced signatures of a pseudo-transition can be traced back to the relevant thermal variations of the entropy and specific heat at moderate values of the magnetic field ($h/J_1 \approx 1$).

\begin{figure}[h!]
\includegraphics[width=0.36\textwidth]{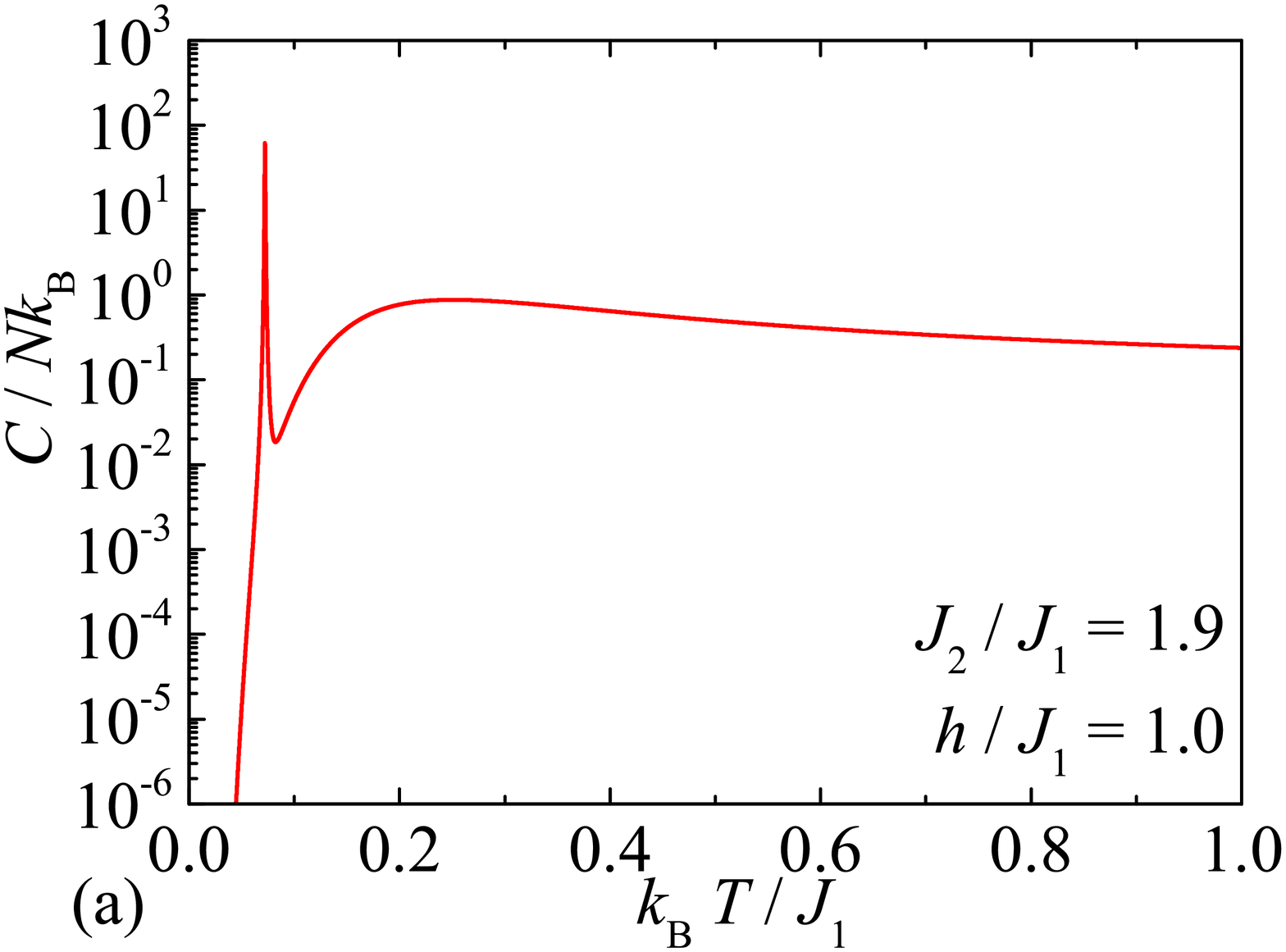}
\hspace{-0.9cm}
\includegraphics[width=0.36\textwidth]{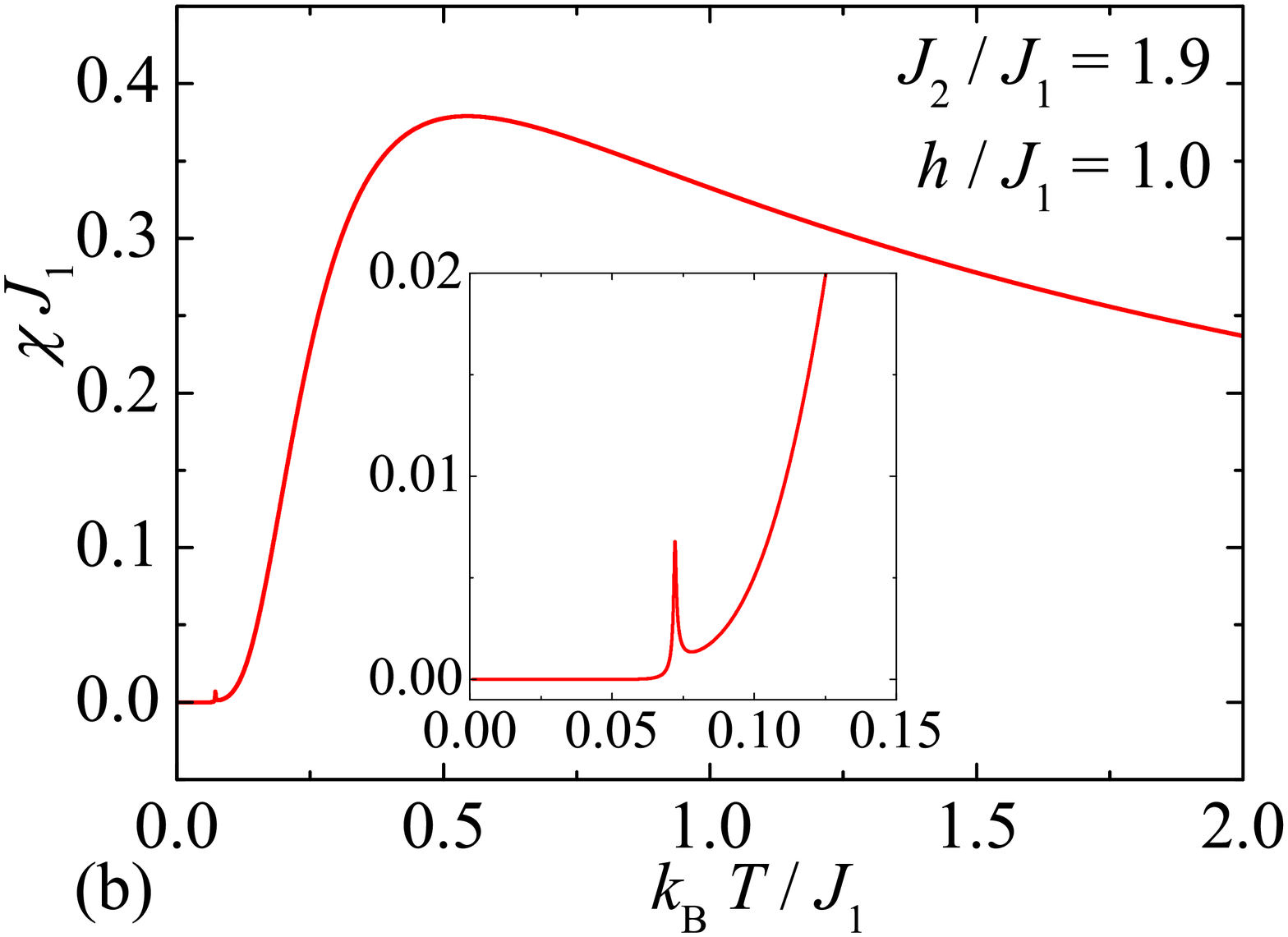}
\hspace{-0.9cm}
\includegraphics[width=0.37\textwidth]{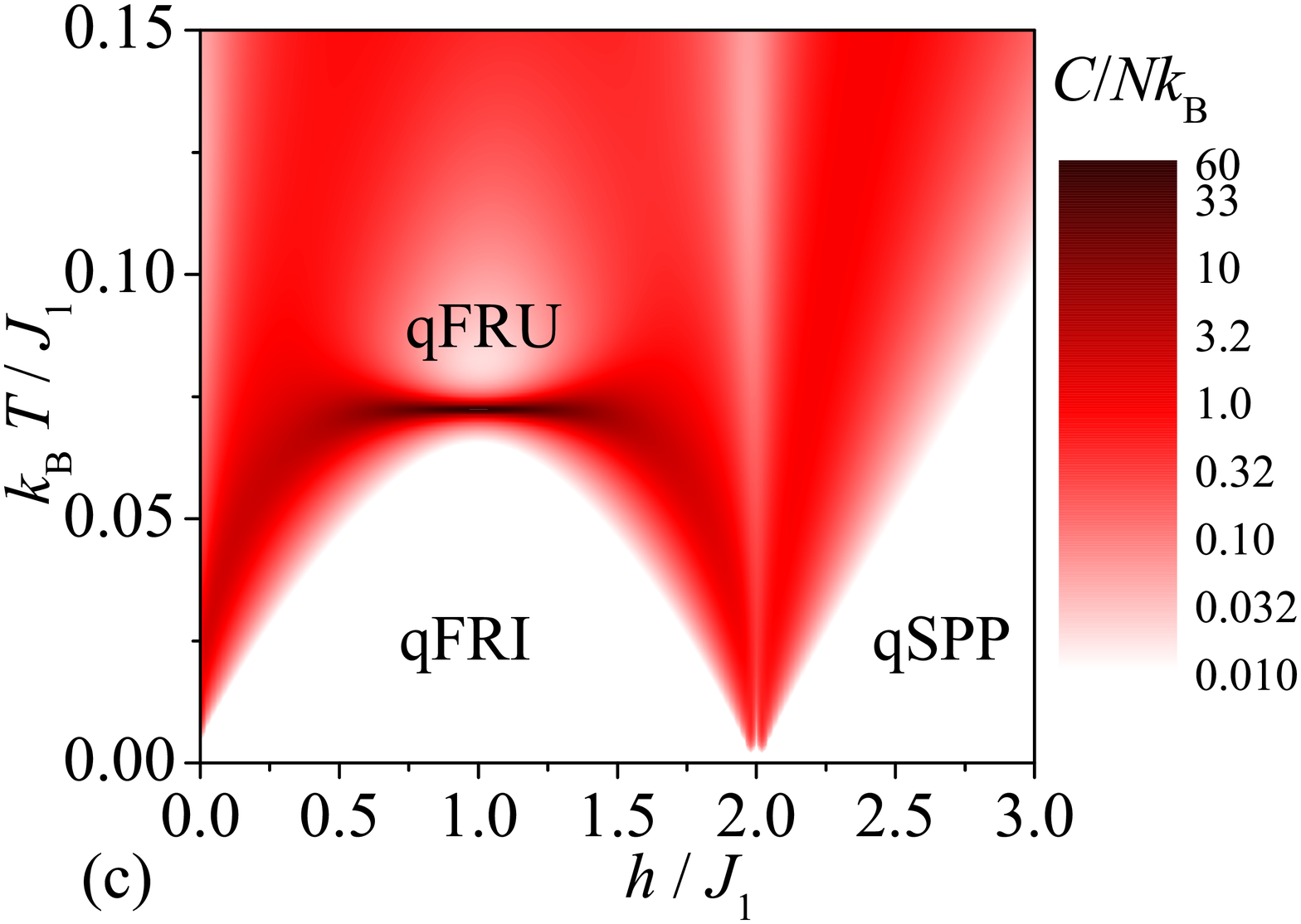}
\vspace{-0.7cm}
\caption{(a)-(b) Temperature variations of the specific heat (semi-logarithmic plot) and susceptibility for the fixed value of the interaction ratio $J_2/J_1 =  1.9$ and the magnetic field $h/J_1 = 1.0$; (c) A density plot of the specific heat in the magnetic field vs. temperature plane for the interaction ratio $J_2/J_1 =  1.9$. The following notation is used for the quasi-phases: 
quasi-ferrimagnetic (qFRI), quasi-frustrated (qFRU) and quasi-saturated-paramagnetic (qSPP) phase.}
\label{fig3}
\end{figure}

Our further attention will be accordingly focused on the moderate values of the magnetic field, which allow a detailed investigation of the pseudo-transition phenomenon in the most authentic manner. For comparison, temperature variations of the specific heat and susceptibility are depicted in Fig. \ref{fig3}(a)-(b) for the fixed value of the interaction ratio $J_2/J_1 =  1.9$ and the magnetic field $h/J_1 = 1.0$. While the specific heat exhibits in the vicinity of the pseudo-critical temperature a substantial rise spread over several orders of magnitude (see the specific-heat plot presented in Fig. \ref{fig3}(a) in a semi-logarithmic scale), the susceptibility shows a qualitatively similar behavior albeit with a much less conspicuous peak at a pseudo-critical temperature [see the inset in Fig. \ref{fig3}(b)]. Hence, it follows that the specific heat represents the best suited quantity for a development of the finite-temperature phase diagram of the spin-1/2 Ising diamond chain in a magnetic field, because it displays the most pronounced temperature-induced changes across a pseudo-critical temperature. As a matter of fact, the density plot of the specific heat shown in Fig. \ref{fig3}(c) in the field-temperature plane clearly allocates all individual quasi-phases. The quasi-ferrimagnetic (qFRI) phase can be located inside of the dome with relatively small values of the specific heat, which is bounded by a quasi-critical fan emanating from the quasi-phase boundary with the quasi-saturated-paramagnetic (qSPP) phase centered at $h/J_1 = 2.0$ and a curvilinear quasi-phase boundary with the quasi-frustrated (qFRU) phase accompanied with a substantial rise of the specific heat on account of its macroscopic degeneracy. 

\section{Conclusions}

In the present work we have furnished an exact evidence that the spin-1/2 Ising diamond chain in a magnetic field may display a remarkable pseudo-transition whenever the model parameters drive the investigated spin chain sufficiently close to a ground-state phase boundary between the non-degenerate ferrimagnetic phase and the frustrated phase with a high macroscopic degeneracy (residual entropy). The emergent pseudo-criticality can be thus attributed to intense thermal excitations from the nondegenerate ferrimagnetic ground state to a highly degenerate manifold of excited states. Among other matters, the pseudo-transition of the spin-1/2 Ising diamond chain manifests itself in an anomalous response of basic thermodynamic quantities, which mimic a temperature-driven phase transition either of a discontinuous (entropy) or continuous (specific heat) nature though there are no true singularities of these quantities at a pseudo-critical temperature. A closer inspection shows that the specific heat and susceptibility follow within a temperature range sufficiently close (but not too close) to a pseudo-critical temperature a power-law dependence, which is characterized by the universal values of the pseudo-critical exponents $\alpha=\alpha'=\gamma=\gamma'=3$ recently reported for a class of 1D lattice-statistical models displaying a pseudo-transition \cite{roj19}. The spin-1/2 Ising diamond chain in a magnetic field thus represents valuable example of 1D lattice-statistical model, which exhibits a pseudo-critical behavior in spite of its fully classical nature.  

\section{Acknowledgement}
This work was financially supported by a grant of the Ministry of Education, Science, Research and Sport of the Slovak Republic under Contract No. VEGA 1/0531/19 and by a grant of the Slovak Research and Development Agency under Contract No. APVV-14-0073.


\end{document}